\documentstyle[amsfonts,epsf,aps]{revtex}
\newcommand{\ee}{\end{equation}}
\newcommand{\be}{\begin{equation}}
\begin{document}
\title{Intercalation and buckling 
instability of DNA linker within locked chromatin fiber}
\author{Jean-Marc Victor, Eli Ben-Ha\"{\i}m, Annick Lesne\footnote{
Email adresses for correspondence: lesne@lptl.jussieu.fr, victor@lptl.jussieu.fr}} 
  \address{
Laboratoire de Physique Th\'eorique
des Liquides, \\
Universit\'e Pierre et Marie Curie,  \\
Case courrier 121, 4 Place Jussieu, 75252
 Paris Cedex 05,
France\\
}
\date{\today}
\maketitle
\vskip 5mm
\begin{abstract}
{\small
The chromatin fiber is a complex of DNA and specific
proteins called histones forming the first structural
level of organization
of eukaryotic chromosomes.
In tightly organized chromatin fibers,
 the short segments of naked  DNA linking
the nucleosomes are strongly end constrained.  
Longitudinal thermal  fluctuations in these linkers
allow intercalative mode of protein binding.
We show that mechanical constraints generated in the 
first stage of the binding process  induce linker DNA buckling;
buckling in turn modifies the
binding energies and activation barriers 
and creates a  force of 
decondensation at the chromatin fiber level.
The unique structure and properties of DNA thus 
yield  a novel physical mechanism of buckling 
instability that might play
a key role in the regulation of gene expression.
}
\end{abstract}
\vskip 20mm

\twocolumn

Notwithstanding its evident biological importance, 
DNA is a fascinating object for physicists
due to its remarkable physical properties.
First, DNA is  a molecular spring
exhibiting  stretch, twist, and bend  elasticities.
These elastic properties have been thoroughly studied
both theoretically\cite{marko} and by single molecule experiments
that led to unexpected results as for instance the structural transitions
observed when a sufficiently strong  pulling force is applied to
 the DNA molecule
\cite{croquette}.
Secondly, DNA is  a  polyelectrolyte and is involved
in numerous electrostatic effects\cite{gelbart}; in particular,
 its negative charge density is enough large to
attract a sheath of counterions almost independent of the salt concentration,
a phenomenon known as
 Manning's condensation\cite{victor}.
Third, the DNA double helix undergoes a denaturation transition
%upon change of ionic strength, of temperature
studied both theoretically \cite{peyrard}\cite{garel}
and experimentally\cite{freifelder}. 
%or mechanical constraints
We here focus on a fourth specific feature of DNA, namely,
intercalation\cite{werner}, that allows 
the binding of planar molecules between adjacent DNA base pairs
(see Fig.~1).

%\vskip 30mm
%

\vspace*{30mm}

\leavevmode
\hspace*{20mm}
\epsfxsize= 80pt
\epsffile[100 150 200  700]{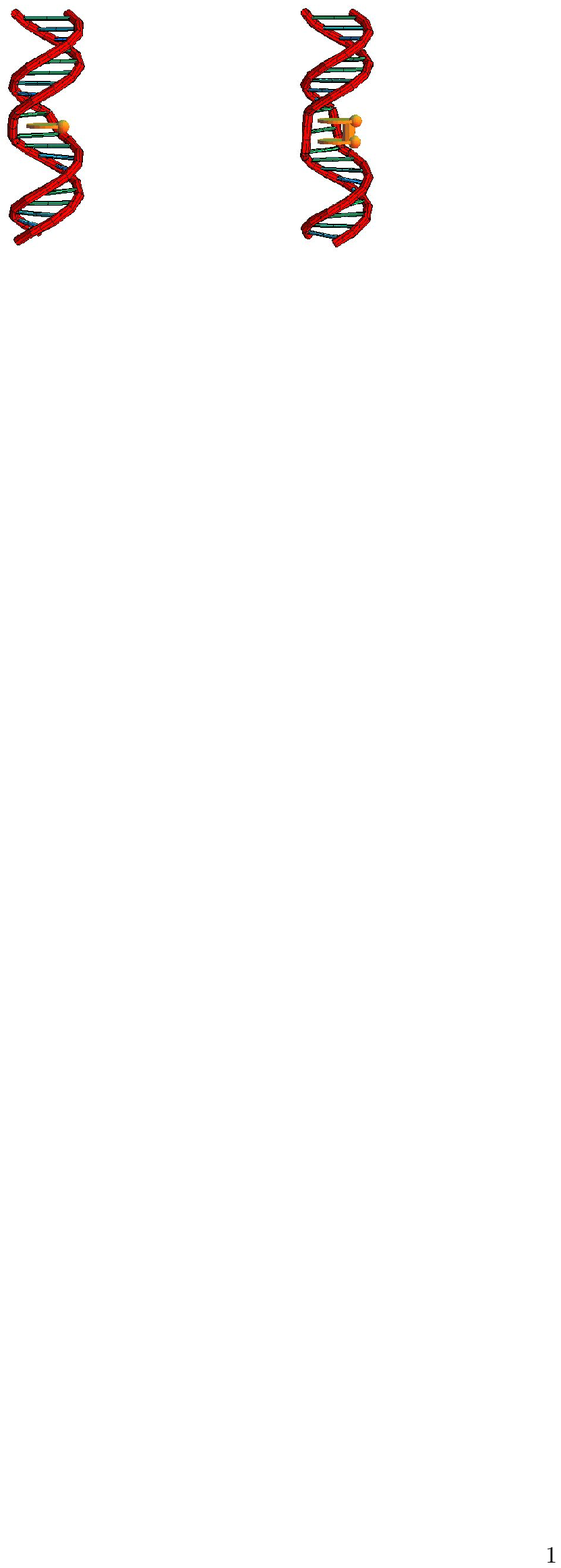}
%\vskip 10mm

\vspace*{-130mm}
\noindent
{\bf Figure 1:} 
Intercalative mode of binding within DNA for a 
mono-intercalator (left) and a bis-intercalator (right).
One (respectively two) domain(s) of the binding protein
comes in between two successive base pairs, inducing
a rise $\Delta l$ and unwinding $-\Delta\tau$.
For instance, ethidium bromide is a mono-intercalator with
$\Delta l\!=\!2$\AA \  and $\Delta\tau\!=\!26^o$.
\vskip 10mm
Quite recently, theoretical \cite{BLV} \cite{schiessel} \cite{langowski}
% +++++++++ ref à ajouter  réordonner.
and experimental
% +++++++++ ref à ajouter
groups \cite{cui-bust} \cite{wang}
started to explore the nanomechanics of chromatin since
free DNA is not the relevant instance of DNA {\it in vivo}.
In the nuclei of plant and animal cells, DNA is actually
organized in a hierarchy of structural levels.
%In all eukaryotic cells, genomic DNA exhibits a hierarchy of structural 
%levels of organization.
The first one is the wrapping of 146 bp of DNA around
histone octamers to form the nucleosomes.
Nucleosomes remain connected
by naked DNA segments --- the linkers --- of length between 10 bp
and 100 bp according to the species and  cell type.
The  nucleosome-dressed DNA  molecule is further organized
%(at least at some stages of the cell cycle)  
into a helical folding
of about 30 nm in diameter that is called ``the chromatin fiber''
\cite{wolffe}.
We recently proposed on mechanical grounds  \cite{BLV}
 that the   chromatin fiber might 
exhibit a columnar packing of nucleosomes
similar to columns observed in colloidal solutions
of mononucleosomes \cite{livolant1}.
%\cite{livolant2};
 It suggests that chromatin might be 
locked into a strongly organized structure, induced  by interactions 
within stacked nucleosomes  and secured by histone tails (Fig.~2).
We call ``locked chromatin'' such a structure
in which the ends of each linker are fixed in space
due to strong three-dimensional positioning of nucleosomes
within the fiber.
A quite  similar structure has been suggested long ago by Worcel 
{\it et al.} \cite{worcel}.
We claim that it provides a plausible  structure
for facultative heterochromatine.

\vskip 15mm
\leavevmode
\hspace*{30mm}
\epsfxsize= 18pt
\epsffile[100 150 200  700]{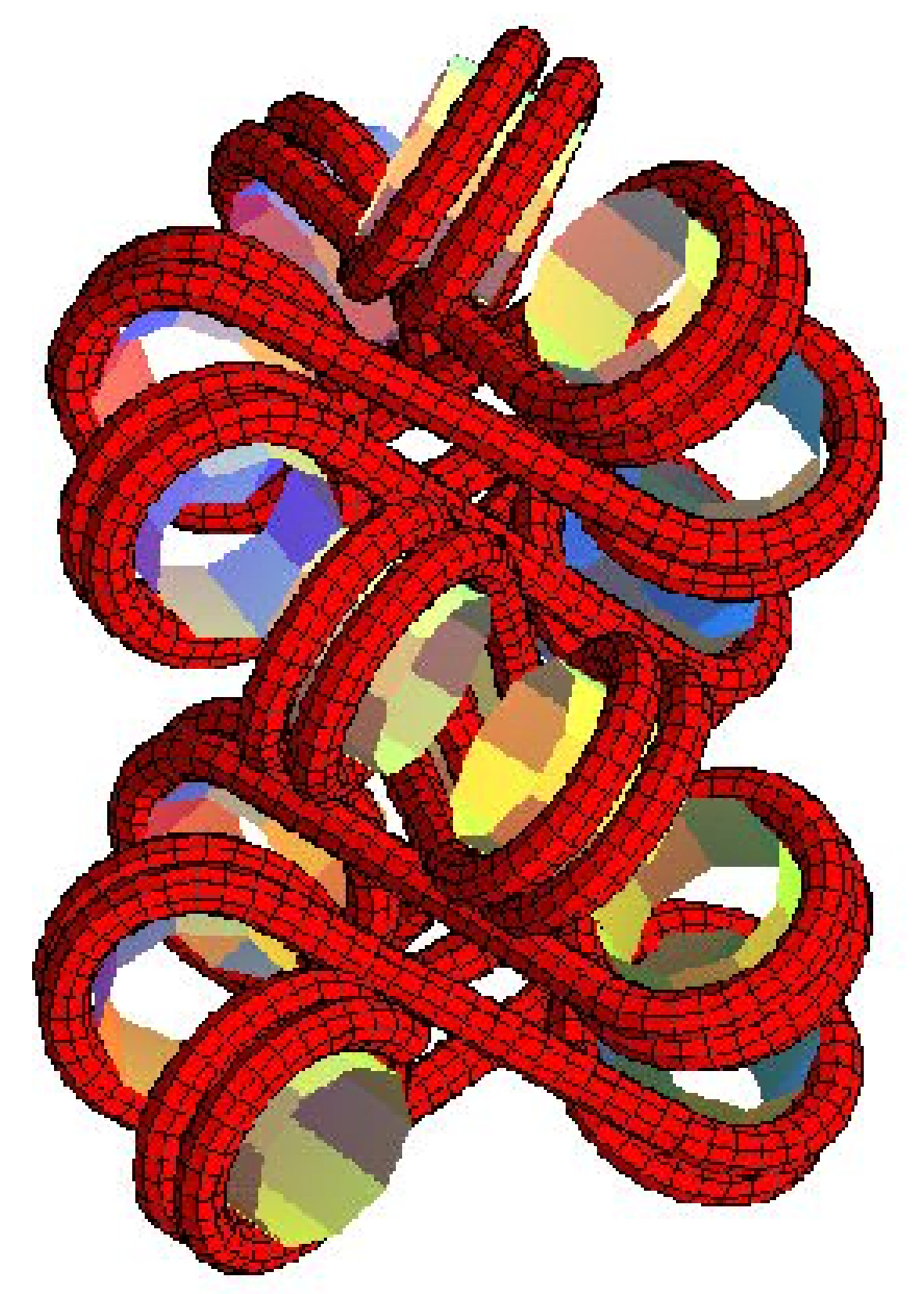}
\vskip 15mm
\noindent
{\bf Figure 2:}
Proposed model of condensed chromatin, locked by interactions between
stacked nucleosomes and presumably also  by histone tails \cite{BLV}.
The chromatin fiber axis is here vertical.
\vskip 10mm

%%%%%%%%%%%%%%%%%%%%%%%%%%%%%%%%%%%%%%%%%%%%%%%%%%%%%%%%%%
\vskip 3mm
In this paper, we propose and describe a different mechanical property occuring
specifically within constrained DNA as encountered in locked chromatin.
It consists of a buckling instability generated by 
longitudinal thermal fluctuations stabilized by intercalation.

Intercalative mode of binding plays an important role {\it in vivo};
for instance the 
%TATA-box-binding protein (
TBP (TATA-box-binding protein), 
which binds 
% is a protein binding
on specific sequences called the TATA boxes and 
plays a seemingly universal role in eukaryotic transcription
initiation, 
%involved in transcription initiation  
is a (multiple) intercalator
\cite{olson}. Intercalation is additionally involved in experimental {\it in vitro}
or {\it in vivo} studies through the use of fluorescent dyes, 
as for instance ethidium bromide \cite{werner}.
In the classical model of intercalation, 
the binding process is decomposed into three steps 
 \cite{chaires1}\cite{chaires2}.
The first step is a thermally activated local opening of DNA, creating a binding site for 
an intercalating molecule.
The activation barrier is of order $\Delta G^0_{conf}\approx 6.5 \;kT$
according to Chaires\cite{chaires1}\cite{chaires2}.
This opening  is achieved through  a local stretching
$\Delta l>0$ of the interbase-pair distance and a local unwinding
$-\Delta \tau$  (Fig.~3).
More precisely, it  corresponds
to a two-state local  conformational transition of the binding site involving
changes in the DNA backbone and  orientation of the bases and sugars\cite{saenger}.
% -- the 
%so-called $\beta$ and 
%$\chi$ angles \cite{saenger}.
The second step is the insertion of intercalator inside the binding site
(hydrophobic transfer process) followed in 
a third step by the formation of molecular interactions, namely,
hydrogen bonds between the intercalator and surrounding base pairs.
The  intercalated site is then slightly smaller in size 
than the preintercalation site required to first accomodate the intercalator.
%(pre-intercalation site).
We underline that the binding site opening  is not induced by the 
presence of intercalator but merely by thermal fluctuations, 
{\it then} stabilized by intercalator insertion:
the mechanism is not an induced fit
but rather a conformational capture, in which intercalation 
captures an ``excited state'' reached spontaneously
by thermal fluctuations \cite{leulliot}.

\vspace*{-28mm}

\leavevmode
\hspace*{20mm}
\epsfxsize= 50pt
\epsffile[100 150 200  700]{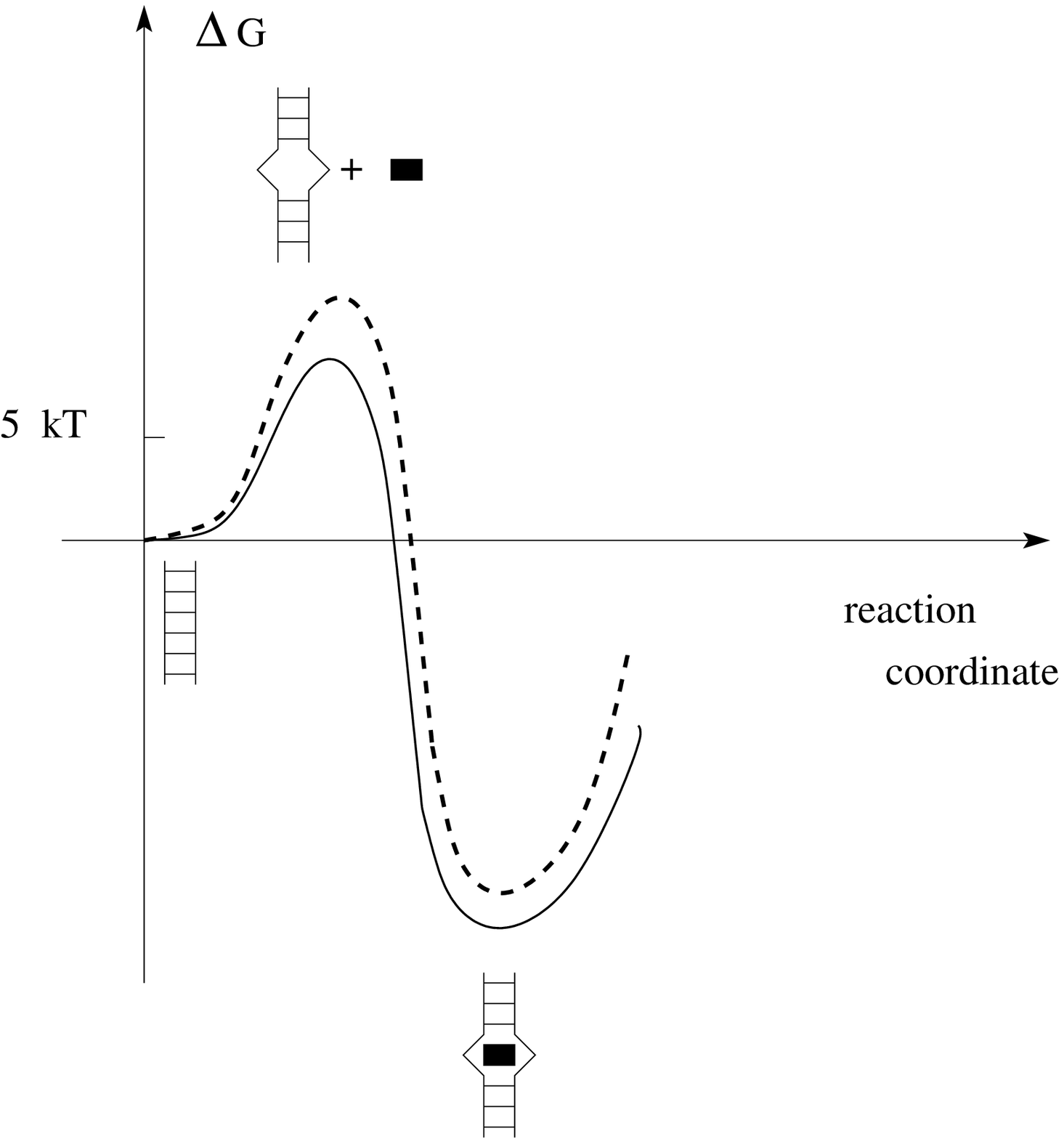}

\vskip 32mm
\noindent
{\bf Figure 3:}
Schematic drawing showing the 
modification of intercalation energy barrier and energy well
coming from mechanical contribution $\Delta\Delta G$.
The units
%magnitudes
 are realistic:
the barrier is of order $7\;kT$; the excess energy is there about $2-3 \;kT$.
The binding energy is of order $-20\;kT$;
the excess energy is there about $1-2 \;kT$
(the intercalated site is smaller than the
``open'' site required in the first step of intercalation,
which reduces $\Delta\Delta G$  into $\Delta\Delta G_{int}$) 
\cite{chaires1}\cite{chaires2}.

%%%%%%%%%%%%%%%%%%%%%%%%%%%%%%%%%%%%%%%%%%%%%%%%%%%%%%%%%%
\vskip 8mm
Let us now consider an intercalation within a linker embedded in a 
locked chromatin fiber, hence with translationally and rotationally
fixed ends. We describe the
linker DNA within a standard continuous model (generalized worm like chain)
as an extensible rod of length $l$ whose mechanical properties are fully described
by  its bending persistence length $A \approx 50$ nm,
its twist persistence length $C\approx 75$ nm
and its 
stretch modulus $\gamma\approx 1200$ pN \cite{marko}.
Since the  distance
between linker ends is fixed, prescribed by the global architecture of
the chromatin fiber, the first step of intercalation
 generates mechanical constraints
in  the remaining part of the linker, namely, 
 a compression  $-\Delta l$ 
and an overtwist $\Delta \tau$, from which follows an 
excess energy $\Delta \Delta G$.
The formation of an intercalation site thus requires to overcome the energy
barrier $\Delta G_{conf}=\Delta G^0_{conf}+\Delta\Delta G$,
%\be\Delta G_{conf}=\Delta G^0_{conf}+\Delta\Delta G
%\hskip 10mm {\rm with}\hskip 10mm
%\ee
with
\be
\Delta \Delta G=\gamma\;{(\Delta l)^2\over 2(l-h)}
+ kTC\;{(\Delta\tau)^2\over 2(l-h)}\ee
where $h$ is the interbase-pair distance. 
We estimate this excess energy with 
$h=3.4$ \AA \ , $\Delta l=3$ \AA \ , $\Delta\tau=30^o$  
(we take these structural values slightly larger than those observed after intercalation,
presuming that the preintercalation site should be larger in size).
We find $\Delta \Delta G\approx 2.4\;kT$
for $l=$ 30 bp (10 nm) and 
$\Delta \Delta G\approx 3.5\;kT$
for $l=$ 20 bp (6.7 nm). 
%\be
%\left\{
%\begin{array}{ll}
%\Delta \Delta G\approx 2.4\;kT&\hskip 10mm{\rm for}\;l=10\;{\rm nm}\;\;(30\;{\rm bp})\\
%&\\
%\Delta \Delta G\approx 3.5\;kT&\hskip 10mm{\rm for}\;l=6.7\;{\rm nm}\;\;(20\;{\rm bp})
%\end{array}
%\right.\ee
%with roughly equal contributions for the twist and stretch terms.
This excess  energy  is  of order of thermal energy
provided  $l$ is not too small; it shows that
thermally activated creation  of an intercalation site  within a constrained linker
is still possible
for $l$ greater than a minimum length $\approx 20$ bp.
We underline that
the thermal fluctuations here invoked are longitudinal fluctuations;
bend fluctuations play a negligible role due to the short length
of the linker compared to the bend persistence length of DNA\cite{odijk}.
Mechanical constraints do not significantly modify the energy
terms involved in the following steps, namely, the free energy cost
associated with the hydrophobic transfer step, 
 the energy associated with modification of counterion surrounding
and the energy of newly established molecular interactions between
 intercalator and DNA \cite{chaires1}.
Note that after intercalation, $\Delta\Delta G$ is to be computed with final
values  of $\Delta l$ and $\Delta\tau$, which gives
a ``mechanical'' correction $\Delta\Delta G_{int}$
to $\Delta G$  that is smaller than  $\Delta\Delta G$.
Mechanical constraints modify both the thermodynamics  of the binding (depth
of the energy well) and the kinetics of the binding
 (height of the activation barrier):
the activation barrier is increased by an amount $\Delta\Delta G$
whereas the binding energy  rises by 
a smaller amount   $\Delta\Delta G_{int}$ (see Fig.~3).

%%%%%%%%%%%%%%%%%%%%%%%%%%%%%%%%%%%%%%%%%%%%%%%%%%%%%%%%%%
\vskip 3mm

Compared to intercalation within free DNA,
the difference lies also
 in the {\it stresses} generated when an intercalation
site opens in an end-constrained linker. 
The force induced by the compression $-\Delta l$ of the
 linker outside the intercalation site writes
\be
F_{int}=\gamma\;{\Delta l\over l-h}
\ee
For $l=30$ bp, i.e. $l=10$ nm, and $\Delta l=3$ \AA \ , 
we obtain $F_{int}=36$ pN.

%%%%%%%%%%%%%%%%%%%%%%%%%%%%%%%%%%%%%%%%%%%%%%%%%%%%%%%%%%

\vskip 3mm

It is well known that above a critical threshold,
 the compression of a  column shifts 
into bending, what is called ``buckling.''
This phenomenon persists up to nanometer scale
\cite{manning}.
In order to determine whether the  force $F_{int}$
will induce, or not,  buckling of the linker,
we compute the critical force $F_c$ 
corresponding to  buckling instability  threshold.
At this stage,
% we have to describe more precisely 
the boundary conditions
at the linker ends, namely, the mode of anchoring of linkers onto the nucleosomes,
have to be described precisely.
Standard computation of material mechanics\cite{timoshenko}
shows that boundary conditions might
be taken into account through a numerical coefficient $\nu$
reflecting the degree of end fixing, e.g. $\nu=1/2$ for clamped ends and
$\nu=1$ for hinged ends;
any value of $\nu\geq 1/2$ is in fact possible for various
elastic joints.
The critical Euler load writes accordingly
\be\label{Fc}
F_c={\pi^2 kTA\over (\nu l)^2},
\ee
where $A$ is the above-mentionned DNA-bending persistence length.
Buckling takes place  if $F_{int}>F_c$  or equivalently if
\be
l>{\pi^2 kTA\over\nu^2\gamma\Delta l}\left(1-{h\over l}\right)\;\approx\; 
{\pi^2 kTA\over\nu^2\gamma\Delta l}\;\equiv \;l_c
\ee
This criterion could have been obtained by comparing
the compression energy $\gamma(\Delta l)^2/2(l-h)$ with the 
critical buckling energy $lF_c^2/2\gamma$.
We get
\be
\left\{
\begin{array}{ll}
{\rm for}\;\;\nu=1&\hskip 2mm F_c\approx 20\;{\rm pN}
\;\;{\rm and}\;\; l_c\approx 
%5.5\;{\rm nm}\;\;(
16\;{\rm bp}\\
&\\
{\rm for}\;\;\nu=1/2&\hskip 2mm F_c\approx 80\;{\rm pN}
\;\;{\rm and}\;\; l_c\approx
% 22\;{\rm nm}\;\;(
65\;{\rm bp}
\end{array}
\right.\ee
hence intercalation 
induces 
buckling in typical linkers (30 bp)
if their ends are hinged but not if they are clamped.
This result shows that for linker length values between 30 and 60 bp,
buckling may be selectively controlled
 by the  anchoring of the linkers at the entry-exit points on 
the nucleosomes. Indeed, according to the biochemical status of the 
nucleosome core, either linker DNA is tightly grafted onto the core,
which corresponds to clamped ends ($\nu=1/2$),
either DNA might unwrap, typically by 5-10 bp on each side,
at negligible energetic cost \cite{wang}\cite{prunell}, which corresponds to 
hinged ends ($\nu=1$) (Fig.~4).

\vspace{-92mm}

\leavevmode
\hspace*{20mm}
\epsfxsize= 50pt
\epsffile[100 150 200  700]{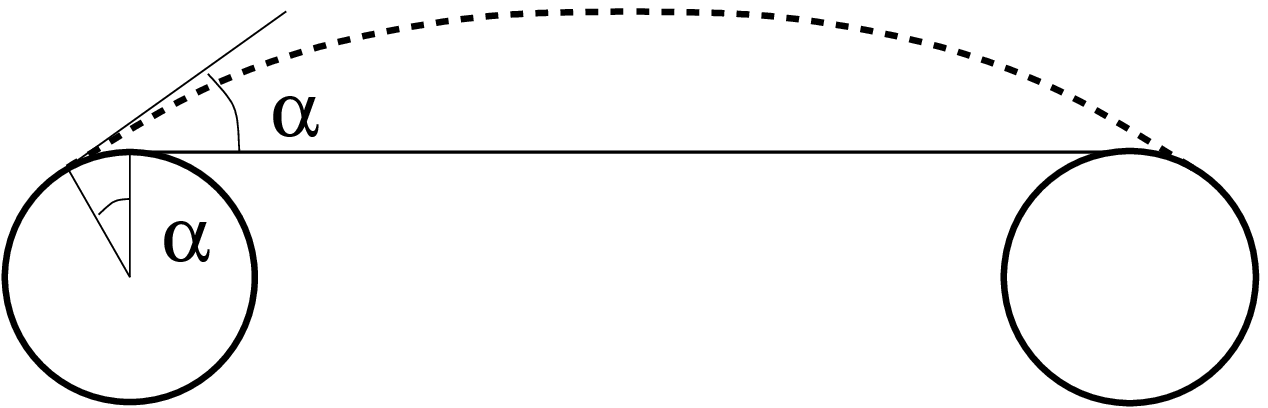}

\vskip 32mm
\noindent
{\bf Figure 4:}
Buckling of the linker is possible
only when the linker ends 
are allowed to unwrap from the nucleosomes.
Unwrapping by 5 bp on each side is enough to accomodate a buckling angle 
$\alpha=20^o$, associated with $\Delta l/l=0.03$;
within these bounds, the situation  is  mechanically equivalent
to hinged ends ($\nu=1$).

\vskip 10mm

Linker buckling in turn modifies the energetics of intercalation.
When buckling occurs, excess energy 
is now the sum of three contributions:
the compression energy 
{\it before} buckling 
$\gamma(\Delta l_b)^2/ 2(l-h)$,  
where $\Delta l_b=F_c(l-h)/\gamma$ is the stretching 
deformation required to reach  the buckling threshold, 
the compression
 energy 
{\it after} buckling 
$F_c(\Delta l -\Delta l_b)$ 
%where $\Delta l$ is the size of the binding site prior intercalation,
and the overall twist energy
$kTC(\Delta\tau)^2/ 2(l-h)$.
It comes
\be
\Delta\Delta G^{buckled}=
%{\pi^2 A\over (\nu l)^2}
F_c\;\left[
\Delta l -{F_c(l-h)\over 2\gamma}\right]+{kTC(\Delta\tau)^2\over 2(l-h)}
 \ee
This expression points out that
once a linker is buckled, the energy cost to stretch it further
is  linear in the length increase and proportional to $1/\nu^2$, hence tunable by 
a modification of the  end status.
%%%%%%% ajout
Buckling thus facilitates the insertion of additional  intercalators,
and perhaps more crucially, it allows the insertion
of multi-intercalators as, for instance,  the above-mentionned TBP.
% (a protein binding
%on specific sequences called the TATA-boxes and 
%playing a seemingly universal role in eukaryotic transcription
%initiation). 

%%%%%%%%%%%%%%%%%%%%%%%%%%%%%%%%%%%%%%%%%%%%%%%%%%%%%%%%%%
\vskip 3mm

At the fiber level,
the buckling force $F_c$ induces stresses,
 that are easy to estimate within our structural modeling of the 
condensed chromatin fiber.
When the two linker ends are anchored in the same way 
on the nucleosome core, 
the force acts along the intercalated linker; it exhibits 
a  component $F_c\cos z$ along the chromatin fiber axis,
where $z$ is the angle between the linkers and this axis
\cite{BLV}.
Buckling also generates a radial shear
% decondensing force,
whose exact expression depends on   the  details
of the chromatin fiber structure. 
%We note that  the proposed locked structure of the fiber
%(Fig.~1)
%requires the  orientational phasing of the nucleosomes, 
%associated with an angle $z$ for which 
%$\cos z$ is around 0.6, close to its maximum value;
%the distance between stacked nucleosomes is correspondingly
%minimal\cite{BLV}.
%T
In the proposed locked structure of the fiber
(Fig.~2), $\cos z$ is around 0.6, close to its maximum value:
this structure is thus  at the same time the more easy to lock thanks
to  interactions between stacked
nucleosomes and the more easy to open thanks to the 
decondensing force generated by intercalation-induced linker buckling.
This two-fold property provides additional support 
of the biological relevance of the proposed structure.
Note that once the linker is buckled,
the force experienced by the linker, hence the force of decondensation does not vary
 significantly with the number of bound intercalator proteins.
%The radial decondensing force equals $2 F_c \sin z$

%%%%%%%%%%%%%%%%%%%%%%%%%%%%%%%%%%%%%%%%%%%%%%%%%%%%%%%%%%

Buckling and decondensing forces are generated by the 
hyperstatic structure of the locked chromatin fiber.
Actually, intercalation induces 
 not only compression but
also torsional strains within the linkers
(which  hence  behave not only as columns 
but also as shafts). 
Twist-induced stresses are yet proportional
to the number of intercalated molecules
and can produce a decondensing force of several piconewtons
along the chromatin fiber axis, 
but buckling, occuring before multiple intercalation, modifies the
picture and bypasses twist effects by turning them into writhe.
We  checked that the resulting contribution  plays a 
secondary role.

%%%%%%%%%%%%%%%%%%%%%%%%%%%%%%%%%%%%%%%%%%%%%%%%%%%%%%%%%%
\vskip 3mm

%From a biological viewpoint,
Taken together, the above results suggest the following scenario 
providing the first physical bases for the
decondensation process:

(i) Thermal fluctuations allow the formation of an intercalation cavity
which induces a compression force in the remaining part of the linker.
According to the mode of linker anchoring  onto the nucleosome core, the
compression force may induce  buckling of the linker.

(ii)  Buckling is
further stabilized by insertion of an intercalator into the cavity.
Moreover, buckling allows the insertion of further intercalators in the same
linker so that the limiting step is indeed buckling.
% the first intercalation;
%Multi-intercalants (e.g. TBP) can be inserted if and only if the linker
%is buckled (
%first insertion allows any further insertion.

(iii) The buckling force generates in turn decondensation forces
at the chromatin fiber level: a
stretching force along the axis of the chromatin fiber as well as
shearing forces between stacked nucleosomes.

%\vskip 1mm
\noindent
This nonlinear decondensation mechanism is irreversible;
chromatin condensation will occur along another pathway, 
 involving electrostatics to induce
 compaction of the fiber (currently under study).

\vskip 3mm

The  physical mechanism here proposed might be of
biological  importance
%since  it might crucially
%participate in 
in  the regulation of transcription. 
Transcription initiation  requires a step of decondensation
of the chromatin  fiber in order to give access to 
transcription machinery\cite{struhl}.
In this paper, we raised the point that
 binding of transcription factors and other 
DNA-binding proteins that monitor the early stages of transcription
should occur within a locked fiber, hence in mechanically
constrained linkers.
Structural modifications
of linker DNA  here play a role through the forces they generate,
without any ATP-consuming mechanism.
This point has yet  been put forward  to account 
for the cooperativity of  protein binding on naked DNA \cite{rudnick}.

Presumably, the chromatin fiber architecture has been deviced 
in the course of evolution not only to efficiently
pack DNA inside the nucleus, but also, as proposed here, 
 to play a mechanically active role in gene expression regulation.

%%%%%%%%%%%%%%%%%%%%%%%%%%%%%%%%%%%%%%%%%%%%%%%%%%%%%%%%%%


\begin{thebibliography}{2000}
\vskip 20mm

\bibitem
{marko} J. Marko, 
{\it Phys. Rev. E} {\bf 57}, 2134 (1998).


\bibitem{croquette}
J.F. Allemand, D. Bensimon, R. Lavery, and V. Croquette,
% Stretched and overwound  DNA forms a Pauling-like
% structure with exposed bases,
{\it Proc. Natl. Acad. Sci. USA} {\bf 95}, 14152 (1998). 



\bibitem
{gelbart} W.M. Gelbart, R.F. Bruinsma, P.A. Pincus,
and V.A. Parsegian, 
% DNA-inspired electrostatics,
{\it Physics Today} {\bf 53} (9) 38, September 2000. 

\bibitem{victor}
For an introductory course, see for example
J.M. Victor, 
% Basic theory of polyelectrolytes, 
p. 291-310 in {\it The physics and chemistry of aqueous
ionic solutions}, ebited by M.C. Bellissent-Funel and G.W. Neilson,
(Reidel, Dordrecht, 1987).

\bibitem{peyrard}
M. Peyrard and A.R.  Bishop,
%Statistical mechanics of a nonlinear model for DNA denaturation.
{\it Phys. Rev. Lett.} {\bf 62}, 2755 (1989).

\bibitem{garel}
T. Garel, C. Monthus, and H. Orland, 
%A simple model for DNA denaturation,
{\it Europhys. Lett.} {\bf 55}, 132 (2001)  

\bibitem{freifelder}
D. Freifelder, {\it Molecular biology}, Jones and Bartlett, Boston (1987).

%\bibitem{essevaz}
%B. Essevaz-Roulet, U. Bockelmann, and F. Heslot,
%Mechanical separation of the complementary strands of DNA
%{\it Proc. Natl. Acad. Sci. USA} {\bf 94}, 11935 (1997).

\bibitem{werner}
M.H. Werner, A.M. Gronenborn, and G.M. Clore,
{\it Science} {\bf 271}, 778 (1996).


\bibitem{BLV} E. Ben Ha\"{\i}m, A. Lesne and J.M. Victor,
{\it Phys. Rev. E} {\bf 64}, 051921 (2001).

\bibitem{schiessel} H. Schiessel, W.M. Gelbart, and R. Bruinsma,
%{\it  Structural and mechanical properties of the two-angle model
%for chromatin}, 
{\it Biophys. J.} {\bf 80}, 1940 (2001).

\bibitem{langowski}
G. Wedemann and J. Langowski, 
%Computer simulation of the 30-nanometer chromatin fiber
{\it Biophys J.} {\bf 82}, 2847 (2002).


\bibitem
{cui-bust} Y. Cui and C. Bustamante,
% Pulling a single chromatin fiber reveals the
%forces that maintain its higher-order structure, 
{\it Proc. Natl. Acad. Sci. USA} {\bf 97}, 127 (2000).

\bibitem{wang} B.D. Brower-Toland, C.L. Smith,
R.C. Yeh, J.T. Lis, C.L. Peterson, and M.D. Wang,
{\it Proc. Natl. Acad. Sci. USA} {\bf 99}, 1960 (2002).

\bibitem{wolffe} A.P. Wolffe, {\sl Chromatin, structure and function},
3rd ed., (Academic Press, San Diego, 1998).



\bibitem
{livolant1} A. Leforestier and F. Livolant, 
{\it Biophysical J.} {\bf 73}, 1771 (1997).



\bibitem
{worcel} A. Worcel, S. Strogatz, and D. Riley,
{\it Proc. Natl. Acad. Sci. USA} {\bf 78}, 1461 (1981).



%\bibitem
%{livolant2} F. Livolant and A. Leforestier, {\it
%Biophysical J.} {\bf 78}, 2716 (2000).









\bibitem{olson} 
K.M. Kosikov, A.A. Gorin, V.B. Zhurkin,
and W.K. Olson, 
{\it J. Mol. Biol.} {\bf 289}, 1301 (1999).



\bibitem{chaires1}
J.B. Chaires,  {\it Biopolymers}
{\bf 44}, 201 (1997).
In this paper, we add an exponent 0 
%$\Delta G^0_{conf}$
to what is called 
$\Delta G_{conf}$ in this paper, namely the energy required to create 
a binding site, to indicate that
 it corresponds to the unconstrained
case (free ends).

\bibitem{chaires2} J.B. Chaires,
 {\it Biochemistry}
{\bf 39}, 8439 (2000).

\bibitem{saenger}
W. Saenger, {\sl Principles of nucleic acid structure},
(Springer, Berlin, 1988), chap. 19.


\bibitem{leulliot}
N. Leulliot and G. Varani,  {\it Biochemistry} {\bf 40}, 7947 (2001).


\bibitem{odijk}
T. Odijk, {\it J. Chem. Phys.} {\bf 108}, 6923 (1998).


\bibitem{manning}
G.S. Manning, 
%Correlation of polymer persistence length with Euler
%buckling fluctuations,
 {\it Phys. Rev. A} {\bf 34}, 4467 (1986).



\bibitem{timoshenko}
S.P. Timoshenko and J.M. Gere, {\it Theory of elastic stability},
(Mac Graw Hill, New York,  1961).



\bibitem{prunell} A. Hamiche and  A. Prunell, 
{\it J. Mol. Biol.}  {\bf 228}, 327 (1992).




\bibitem{struhl}
K. Struhl, D. Kadosh, M. Keaveney, L. Kuras, and Z. Moqtaderi, 
{\it Cold Spring Harbor Symp. Quant. Biol.} {\bf 63}, 413 (1998).

\bibitem{rudnick}
J. Rudnick and R. Bruinsma, 
{\it Biophys. J.} {\bf 76}, 1725 (1999).


\end{thebibliography}
\end{document}